%% file: MAIN.tex
\documentclass[%
 reprint,
nofootinbib,
 amsmath,amssymb,
 aps,
]{revtex4-2}

\usepackage{graphicx}
\usepackage{dcolumn}
\usepackage{booktabs}
\usepackage{wasysym}
\usepackage[normalem]{ulem}
\usepackage{bm}

\usepackage{color}
\usepackage{xcolor}

\usepackage{xcolor}
\usepackage{xcolor}

\newcommand{\INFNRM}{INFN, Sezione di Roma, Piazzale Aldo Moro, 2, I-00185 Roma, Italy}

\newcommand{\Sapienza}{Physics Department, Universit\`a di Roma “Sapienza”, Piazzale Aldo Moro, 2, I-00185 Roma, Italy}
\newcommand{\Balea}{IAC3–IEEC, Universitat de les Illes Balears, Carretera de Valldemossa km 7,5, E-07122 Palma de Mallorca, Spain}

\usepackage{tikz}

\begin{document}

\preprint{APS/123-QED}

\title{Harnessing the potential of PyStoch: Detecting continuous gravitational waves from interesting supernova remnant targets}


\author{
\textsuperscript{§}Claudio Salvadore\textsuperscript{1,2},
\textsuperscript{*}Iuri La Rosa\textsuperscript{3}, 
\textsuperscript{‡}Paola Leaci\textsuperscript{1,2}, 
Francesco 
Amicucci\textsuperscript{1,2}, 
Pia Astone\textsuperscript{2},
Sabrina D'Antonio\textsuperscript{2},
Luca D'Onofrio\textsuperscript{2},
Cristiano Palomba\textsuperscript{2},
Lorenzo Pierini\textsuperscript{2},
Francesco Safai Tehrani\textsuperscript{2}
}
\affiliation{\textsuperscript{1} \Sapienza}
\affiliation{\textsuperscript{2} \INFNRM}
\affiliation{\textsuperscript{3} \Balea}
\input{sections/abstract.tex}
\noindent
\textbf{Journal reference:} Phys. Rev. D 112, 083051 (2025). \\
\textbf{DOI:} \href{https://doi.org/10.1103/xydb-k4vw}{https://doi.org/10.1103/xydb-k4vw} \\
\textbf{Comments:} Published version includes minor editorial changes.
\maketitle
\begin{flushleft}
\textsuperscript{§}\texttt{Contact author: claudio.salvadore@roma1.infn.it}\\
\textsuperscript{*}\texttt{Contact author: iuri.larosa@uib.eu} \\
\textsuperscript{‡}\texttt{Contact author: paola.leaci@roma1.infn.it}
\end{flushleft}

\input{sections/first_part}

\input{sections/second_part}

\bibliography{sections/biblio}

\end{document}

%% file: sections/abstract.tex
\begin{abstract}

Continuous gravitational waves (CWs) from nonaxisymmetric neutron stars (NSs) are key targets for the Advanced LIGO-Virgo-KAGRA detectors. While no CW signals have been detected so far, stringent upper limits on the CW strain amplitude have been established.
Detecting CWs is challenging due to their weak amplitude and high computational demands, especially with poorly constrained source parameters. Stochastic gravitational-wave background (SGWB) searches using cross-correlation techniques can identify unresolved astrophysical sources, including CWs, at lower computational cost, albeit with reduced sensitivity. This motivates a hybrid approach where SGWB algorithms act as a first-pass filter to identify CW candidates for follow-up with dedicated CW pipelines.

We evaluated the discovery potential of the SGWB analysis tool PyStoch for detecting CWs, using simulated signals from spinning down NSs. We then applied the method to data from the third LIGO-Virgo-KAGRA observing run (O3), covering the ($20-1726$) Hz frequency band, and targeting four supernova remnants: Vela Jr., G347.3-0.5, Cassiopeia A, and the NS associated with the 1987A supernova remnant.
If necessary, significant candidates are followed up using the 5-vector Resampling and band-sampled data frequency-Hough techniques.
However, since no interesting candidates were identified in the real O3 analysis, we set 95\% confidence-level upper limits on the CW strain amplitude $h_0$. The most stringent limit was obtained for Cassiopeia A, and is $h_0 = 1.13 \times 10^{-25}$ at $201.57$ Hz with a frequency resolution of $1/32$ Hz. As for the other targets, the best upper limits have been set with the same frequency resolution, and correspond to $h_0 = 1.20 \times 10^{-25} $ at $202.16$ Hz for G347.3-0.5, $1.20 \times 10^{-25}$ at $217.81$ Hz for Vela Jr., and $1.47 \times 10^{-25}$ at $186.41$ Hz for the NS in the 1987A supernova remnant. \\

\end{abstract}

%% file: sections/first_part.tex
\section{\label{sec:intro}Introduction}
Continuous gravitational waves (CWs) are a crucial class of signals anticipated to be detected by the advanced LIGO-Virgo-KAGRA detectors. These signals, emitted by rapidly spinning neutron stars (NSs) with structural asymmetries \cite{FIRST,carl}, present one of the most fascinating challenges in modern gravitational-wave (GW) astrophysics. The search for CWs from supernova remnants such as supernova 1987A (SN1987A), Vela Jr., G347.3-0.5 (G347), and Cassiopeia A (CasA)  is motivated by the potential presence of NSs within these remnants \cite{FIRST,Wang,wette,EH,EH2,Vela,G347,CasA,panagia,Lin}. Detecting such signals would offer profound insights into the internal structure of NSs, their equation of state, and the underlying mechanisms driving their formation and evolution. Furthermore, SN1987A holds particular significance, being one of the closest and most extensively studied supernovae, presenting a rare opportunity to explore the physics that follows a supernova explosion. Although previous studies using LIGO and other detectors have searched for CWs from these sources \cite{wette,Wang,EH,EH2,owen,OwenLin,papa,SNR1,SNR2,SNR3,SNR4, Lin}, no definitive detections have yet been made. Nevertheless, with progressively more sensitive instruments such as the new advanced LIGO-Virgo-KAGRA detectors \cite{Adv} and the future third-generation detectors, i.e. the Einstein Telescope \cite{ET} or the Cosmic Explorer \cite{CE} there is great potential to observe new and groundbreaking astrophysical discoveries.

CW searches are typically divided into three categories: targeted \cite{Targ}, directional \cite{OwenLin,Wang,EH,EH2,dirFH,SNR1,SNR2,SNR3,SNR4,owen,Lin}, and all-sky \cite{allskyO3} searches.
These categories are distinguished primarily by the volume of parameter space they cover, which directly correlates with the computational complexity of the data analysis. This paper focuses on directed searches towards specific supernova remnants, such as Vela Jr., G347, CasA, and SN1987A, where the source position is known, but the frequency and its time evolution remain uncertain, making the analysis more computationally intensive.

The stochastic gravitational-wave background (SGWB) results from the superposition of GW signals from a broad spectrum of astrophysical and cosmological sources. Recent studies \cite{O1-aniso,O2-aniso,O3-aniso,O3-asaf,Messenger,iuriPhd} have shown that directional SGWB searches, while less sensitive, can also detect CW signals. They offer a significant advantage in terms of computational efficiency, requiring less resources compared to traditional, dedicated CW data-analysis pipelines. This approach holds great promise for identifying CW signals, particularly from sources with poorly constrained parameters.

This paper introduces a hybrid approach where PyStoch, i.e. a Python-based tool for SGWB mapping via the radiometer method \cite{Bruce,mitra}, is used to efficiently identify potential CW candidates, which are subsequently scrutinized with dedicated CW data analysis pipelines.

The article is structured as follows. Sec. \ref{sec:model} introduces the fundamental theoretical concepts behind CW signals. Sec. \ref{sec:radiometer} presents the GW radiometer technique, which cross-correlates data from two detectors and enables directed CW searches with PyStoch. Sec. \ref{sec:method} describes the search methodology, including the implementation of PyStoch, its evaluation on simulated data, and the candidate selection process. It also introduces the 5-vector (5-vec) resampling \cite{Francesco} and band-sampled data (BSD) frequency-Hough \cite{FH} techniques used for follow-up. Sec. \ref{sec:search} reports the application of the method to O3 LIGO-Virgo-KAGRA data (April 1, 2019 – March 27, 2020 \cite{o3}). No significant candidates are confirmed, and we therefore set 95\% confidence-level upper limits (ULs) on the CW strain amplitude. Finally, Sec. \ref{sec:outro} evaluates the methodology and discusses the results of the O3 searches.

\section{Signal model}
\label{sec:model}

Typically, astrophysical sources like NSs are compact objects with a mass of approximately 1.5 $M_{\astrosun}$ and a radius of around 10 km, resulting from the core-collapse supernova of a star exceeding  $8 M_{\astrosun}$ in its final evolutionary phase \cite{sing}. 

CWs are long-duration (months or years) signals modeled as quasisinusoidal waveforms, likely emitted from nonaxisymmetric, rapidly rotating NSs. Considering an object, steadily spinning around one of its principal inertia axes, the expected CW strain amplitude at the detector is given by \cite{carl}
\begin{equation}\label{ns_strain}
    h_0(t) = \dfrac{16 \pi^2 G}{c^4} \dfrac{\epsilon I f_{\text{GW}}^2(t)}{r},
\end{equation}
where $I \sim 10^{38}$ kg m$^2$ is the moment of inertia of the NS with respect to the rotational axis, G is the universal gravitation constant, $c$ is the speed of light, $\epsilon$ is the NS ellipticity, a measure of its spherical deformation \cite{Ferrari}, $r$ is the distance to the source and $f_{\text{GW}}(t)$ is the CW emitted signal frequency. This function slowly decreases with time due to the rotational energy loss of the star, as a consequence of both electromagnetic and GW radiation. This is the so-called spin-down effect, which can be well described by a Taylor series expansion \cite{picci}:
\begin{equation}\label{spindown}
    f_{\text{GW}} (t) = f_0 + \Dot{f}_0 (t-t_0) + \frac{1}{2} \ddot{f}_0 (t-t_0)^2 + \ldots < f_0,
\end{equation}
where the frequency time derivatives represent the spin-down parameters and $t_0$ is the signal reference time. 

Given Eq \ref{spindown}, during an observation time $T_{\text{obs}}$, the amplitude spectral density of a spinning-down signal will be distributed over a frequency band $\Delta f_{\text{GW}}$, whose width is determined by the relation:
\begin{equation}\label{Deltaf}
    \Delta f_{\text{GW}}  = \Dot{f}_0 T_{\text{obs}} + \frac{1}{2} \Ddot{f}_0  T_{\text{obs}}^2 + \ldots
\end{equation}

In many cases, limited or no information is available about NSs located within supernova remnants \cite{Catalog}, leaving us primarily with estimates of their distance and age $t_{\text{age}}$. In these cases, though, the CW amplitude can be estimated directly using the so-called braking index $n = f_0 \Ddot{f}_0/\Dot{f}_0^2$~\cite{owen,carl}, i.e.,
\begin{equation}\label{strain}
    h_0^{\text{age}} = \frac{2}{\mu \, r}\sqrt{\frac{5GI}{2(n-1)t_{\text{age}}c^3}}.
\end{equation}

The value of the braking index ranges from 2 to 7, according to different energy loss mechanisms\footnote{It takes value $n=3$ for magnetic dipole emission, 5 for quadrupolar GWs, and 7 for r-modes.}, while the parameter $\mu$ represents the ratio between the GW frequency and the star spin frequency\footnote{For r-mode emission, $\mu = 4/3$, while for mass-quadrupole GW emission (``mountain" mechanism), $\mu = 2$~\cite{owen}.}.

As for the amplitude, the spin-down parameters can be estimated from the braking index and the age of the NS. However, an assumption on the signal frequency is required to estimate the spin-down parameters \cite{owen,carl}:
\begin{equation}\label{fdot}
\begin{array}{cc}
\Dot{f}_0 = -\dfrac{f_0}{t_{\text{age}}(n-1)}; & \;\;\Ddot{f}_0 = \dfrac{n\Dot{f}_0^2}{f_0}.\\
\end{array}
\end{equation}

Hence, the expected frequency distribution due to the spin-down effect described in Eq. (\ref{Deltaf}) becomes
\begin{equation}\label{Deltaf2}
    \Delta f_{\text{GW}}  = \underbrace{-\dfrac{f_0}{t_{\text{age}}(n-1)}}_{\dot{f}_0}  T_{\text{obs}} + \frac{1}{2} \underbrace{\dfrac{n}{f_0} \biggl ( \dfrac{f_0}{t_{\text{age}}(n-1)} \biggr )^2}_{\Ddot{f}_0}  T_{\text{obs}}^2,
\end{equation}
where the only free parameter is the signal frequency $f_0$.

A CW signal reaches the detector with its frequency modulated by Doppler shifts induced by the Earth’s orbital and rotational motion. The received signal frequency $f(t)$ is then related to $f_{\text{GW}}(t)$ by \cite{carl}
\begin{equation}\label{Doppler}
    f(t) = f_{\text{GW}}(t) \left ( 1 + \dfrac{\mathbf{v}(t)\cdot \mathbf{\hat{n}}}{c} \right ), 
\end{equation}
where $ \mathbf{v} = \mathbf{v}_{orb} + \mathbf{v}_{rot}$ is the detector velocity, sum of the Earth’s orbital and rotational velocity, while $\mathbf{\hat{n}}$ is the unit vector pointing to the source position, both expressed in the solar system barycenter reference frame.

As a consequence, the combined orbital and rotational motions of the Earth introduce an additional frequency spreading in the received signal:
\begin{equation}
\Delta f_{\text{Doppler}}  \sim 10^{-4}f_0 \text{cos}\beta,    
\end{equation}
with $\beta$ denoting the source ecliptic latitude. The importance of this effect depends on the source parameters: for sources located near the ecliptic poles (such as SN1987A), it is generally negligible at any frequency; at lower frequencies and for typical spin-down values, spin-down spreading dominates at all latitudes; whereas at higher frequencies and lower latitudes, the Doppler contribution can become comparable. The standard approach, then, is to take into account the two effects defining a total frequency spread
\begin{equation}\label{TOTspread}
    \Delta f  \approx \Delta f_{\text{GW}}+\Delta f_{\text{Doppler}},
\end{equation}
where the highest search frequency is used to quantify the Doppler contribution (given the source latitude).

\section{GW radiometer}
\label{sec:radiometer}

Radiometric techniques are employed to create sky maps of anisotropies in the SGWB by cross-correlating data from pairs of detectors \cite{PyStoch}. The GW radiometer algorithm accounts for the delay in the time of arrival of GW signals at detectors located at different positions. Fixing a direction in the sky, this delay varies as the baseline orientation changes due to Earth's motion \cite{PyStoch}. When time-delayed data from two detectors are cross-correlated, potential GW signals arriving from the given direction interfere constructively, while the noise contributions do not. 

For persistent signals, this results in a mismatch in phase evolution, which can be corrected by properly cross-correlating the frequency-domain data from the two detectors $i=1, 2$. The correction factor is a filter composed of the signal spectral template function $H$, the power spectral density estimates for each detector $P_{i}$, and the so-called overlap reduction function, which depends on the source sky position, and is given by 

\begin{equation}
\gamma_f (\Vec{\Omega}) =  \frac{1}{2} \sum _{A=\times,+} F^{A}_{1,f} F^{A}_{2,f} e^{i2\pi f \Vec{\Omega} \cdot \Delta x /c},
\end{equation} 

where $\Vec{\Omega}$ is the unit vector pointing to the CW source and $F^{A}_{i,f}$ are the antenna patterns for the two GW polarizations $A=\times,+$ and the two detectors $i=1,2$.

Over extended observation periods, the signal cross-correlation grows faster than the noise variance, making the detection statistic progressively more significant. 

Since the baseline orientation relative to the target evolves over time, and the noise nonstationarities, the received time series from the two detectors are split into short segments \cite{Bruce}. These two effects can be safely neglected if the coherence time $T_{\mathrm{coh}}$ is shorter than 200 seconds\footnote{We keep the minimum requirement that $T_{\mathrm{coh}}$ must be significantly longer than the light travel time between the detectors, which is approximately 30 ms for LIGO Hanford and LIGO Livingston.}~\cite{mitra}.

Because of this, in the GW radiometer pipeline, given a time series $s_i$, recorded by the detector $i=\{1, 2\}$, the data is sampled with a coherence time $T_{\text{coh}} = 192$ seconds \cite{mitra}. The detection statistics $Y$ is then computed in the frequency domain with a semicoherent approach, cross-correlating the time series Fourier transforms $\tilde{s}_{i}$ for each segment $t$ and then integrated. In the process, the data are sampled to obtain a value of $Y$ for each frequency bin with a certain resolution $\delta f$. Because of the directional dependence of $\gamma$, the result depends on the targeted sky direction, i.e.\cite{O1-aniso,O2-aniso,O3-aniso,Bruce}:
\begin{equation}\label{DS}
    Y_{f} = \dfrac{4H_f}{ \sigma_{Y_f}^2 T_{\text{coh}}} \sum_t \dfrac{\gamma_{ft}}{P_{1_{ft}}P_{2_{ft}}} \tilde{s}^*_{1_{ft}} \tilde{s}_{2_{ft}},
\end{equation}
where $ \sigma_{Y}^2$ is the variance of the cross-correlation statistics \cite{UL}, i.e.:

\begin{equation}
         \sigma^2_Y = \frac{2P_1 P_2}{T_{coh}^2 \sum_t (F^+_{1t} F^+_{2t} + F^\times_{1t} F^\times_{2t})^2}.
\end{equation}

A temporal symmetry was observed in the evolution of the detection statistics [Eq. (\ref{DS})] as a function of both frequency and acquisition time \cite{PyStoch}. This symmetry enables the implementation of the so-called \textit{folding procedure}, which compresses months of data into a single sidereal day. The compactness of the folded data is then exploited by PyStoch~\cite{PyStoch}, which processes the folded (even one-year-long) data within minutes. During this processing, the cross-correlation statistics and the corresponding variance for each frequency bin (with width defaulted to 1/32 Hz) are calculated for the desired sky direction \cite{PyStoch}. Finally, using these results, the signal-to-noise ratio (SNR) for each frequency bin can be computed as \cite{UL}

\begin{equation}\label{snr}
    \text{SNR}_f = \dfrac{\text{Y}_f}{\sigma_{Y_f}}.
\end{equation}


\subsection{Frequency bin combination strategy}\label{SubA}

An essential step in the analysis is the combination of adjacent frequency bins, aimed at enhancing sensitivity to signals with time-varying frequency evolution \cite{O1-aniso,O2-aniso,O3-aniso}. In the standard SGWB searches, a default frequency resolution $\delta f _{\text{def}} = 1/32$ Hz builds, as stated before, a set of contiguous frequency bins, with associated detection statistics $Y_f$ and standard deviation $\sigma_{Y_f}$, yielding the SNR in Eq. (\ref{snr}). To probe broader-band features or reduce statistical fluctuations, bins are combined using a sliding-window approach. 

For each central bin, the combination includes $N$ full bins plus half a bin on either side to take into account signal frequency variations. The combined frequency bin width $\delta f_{\text{comb}}$ is defined as
\begin{equation} \label{comb}
   \delta f_{\text{comb}} = (2N + 1) \delta f  _{\text{def}},
\end{equation}
and the value taken by $N$ ensures that $\delta f_{\text{comb}}$ matches the frequency distribution $\Delta f$ specified in Eq. (\ref{TOTspread}); the default case, corresponding to no bin combination (i.e., $\delta f_{\text{comb}}=\delta f_{\text{def}}$), is denoted by $N=0$.

The SNR for the combined bin is computed as:
\begin{equation}
    \mathrm{SNR}_{\text{comb}} = \frac{\sum_i Y_i}{\sum_i \sigma_{Y_i}},
\end{equation}
where the sum extends over all bins within the combination window centered at each original frequency bin.


It is important to note that this procedure does not reduce the number of evaluated frequency points. The combination window is shifted one bin at a time across the frequency band, and for each shift, an SNR value is computed. This ensures that an SNR value is produced for every original frequency bin, resulting in the same number of data points in the SNR versus frequency array, regardless of the combination width $N$.
We clarify that the combination of neighboring frequency bins is not meant to redefine the intrinsic frequency resolution of the search. Instead, it acts as a running average that enhances the detectability of coherent signals across broader frequency regions, while preserving the underlying nominal binning. In practice, the analysis is performed simultaneously over multiple values of $N$, chosen according to different NS spin-down parameters. This multiscale approach allows us to remain sensitive to signals that may not align exactly with a single frequency bin, while retaining the original resolution when needed. Importantly, bin combination plays a dual role: it is used both to estimate statistical significance (e.g., $p$ values) and as a central ingredient of candidate selection. This dual use, together with the systematic study of how bin combination relates to spin-down parameters (see Sec.~\ref{sec:selection}), represents a key innovation of the pipeline.

\subsection{Detection significance and ULs}
\label{subsec: uls}

The statistical significance of the results is assessed using $p$ values, which indicate the probability that an observed SNR could result from random noise fluctuations \cite{O1-aniso,O2-aniso,O3-aniso} . 

They are derived through Monte Carlo simulations based on Gaussian realizations that accurately reflect the noise properties of the dataset. A large number of simulated SNR distributions are generated, computing the detection statistic $Y_f$ for each instance and for every frequency bin. The values are drawn from Gaussian distributions whose widths are set by the $\sigma_{Y_f}$, obtained directly from the data. Subsequently, they are processed through the same bin combination strategy described in Sec. \ref{SubA}. The maximum SNR from each realization is recorded, and the resulting ensemble is used to construct an empirical mapping between SNR values and their corresponding $p$ values via linear interpolation.
We clarify that in stochastic searches the cross-correlation SNR can take both positive and negative values. In our analysis, only positive SNR excursions are regarded as potentially significant, since they correspond to excess coherence consistent with a physical signal. Frequency bins are considered significant when they yield $p$ values $\leq 10\%$, which in practice corresponds to SNRs $\gtrsim 4.5$. Negative SNR values are not treated as candidates. In the absence of a significant detection, ULs on the CW strain amplitude $h_0$ are placed. The computation of ULs is performed within a Bayesian framework \cite{O1-aniso,O2-aniso,O3-aniso}, incorporating prior distributions over relevant source parameters such as the inclination angle, polarization angle, and calibration uncertainty. The final result is the marginalized posterior distribution for $h_0$, from which the UL at a given confidence level (e.g., 95\% in our case) is extracted.

Because of the computational cost associated with evaluating the full marginalization for each frequency bin, an interpolation-based approach is adopted. ULs are precomputed for a range of representative SNR values and both circular and generic polarizations. The ratio between these cases, which depends only on the SNR, allows for a rapid estimation of the marginalized ULs across the entire frequency band.
 Further technical and mathematical details
 of the $p$ value estimation and UL computation can be found
 in Supplemental Material \cite{O1-supp}.

%% file: sections/second_part.tex
\section{Search Method}
\label{sec:method}

In this section, we outline the methodology, based on the GW radiometer pipeline~\cite{PyStoch}, tuned to search for NSs in supernova remnants. Specifically, we present tests conducted on simulated data to assess the performance of PyStoch in detecting this class of localized CW signals. Additionally, we describe the approach used to identify software injections in the simulated data, which will be applied during the O3 real data search. 

\begin{figure}
    \centering
    \includegraphics[width=1\linewidth]{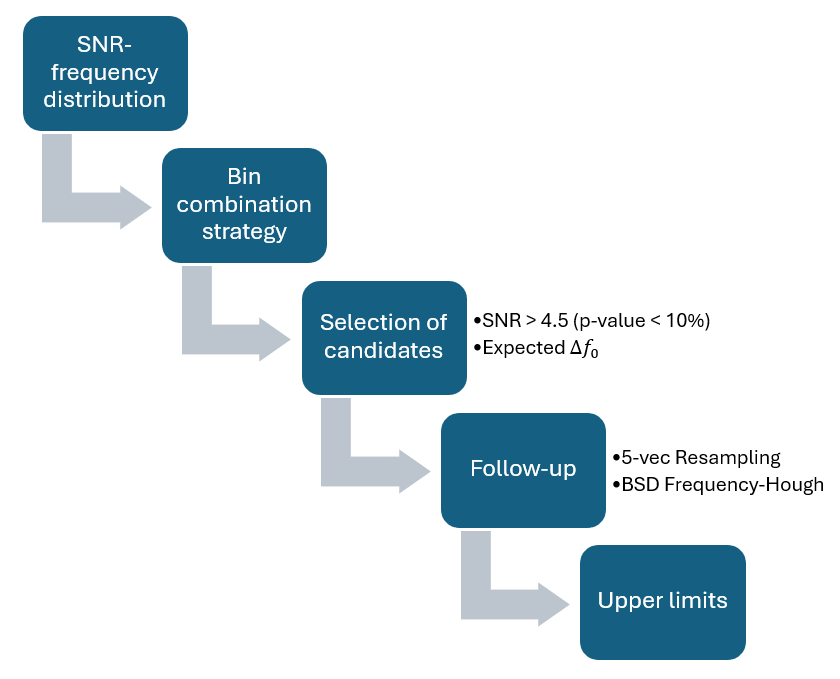}
    \caption{Flowchart of the PyStoch search for CWs targeting SN1987a, Vela Jr., G347 and CasA. First, PyStoch processes cross-correlated and folded data from LIGO Hanford and Livingston, producing narrow-band SNR maps across frequencies from 20 to 1726 Hz, with a default resolution of 1/32 Hz. Next, using the frequency bin combination strategy [Eq. (\ref{comb})], candidates are identified when the following conditions are simultaneously satisfied: SNR $>$ 4.5, i.e. $p$ value $<$ 10\% (Sec. \ref{subsec: uls}), and a frequency distribution $\Delta f$ consistent with theoretical expectations (Sec. \ref{sec:model}).
    Promising candidates undergo further scrutiny via 5-vec Resampling~\cite{Francesco}) and BSD frequency-Hough methods~\cite{FH}.
    In the absence of confirmed candidates, 95\% confidence-level ULs on the strain amplitude are computed (Sec. \ref{subsec: uls}).}
    \label{fig:FC}
\end{figure}

This approach involves combining multiple frequency bins, as described in Sec. \ref{SubA}.
We then outline the candidate selection process (Sec. \ref{sec:selection}), which involves several steps to refine significant candidates:
starting from the SNR-frequency distribution produced by PyStoch (first block in Fig. \ref{fig:FC}), we apply the frequency bin combination strategy (second block in Fig. \ref{fig:FC}). Candidates are then selected when the following conditions are simultaneously satisfied: SNR$>4.5$, i.e. $p$ value$<$ 10\% (see Sec. \ref{subsec: uls}), and a frequency evolution consistent with theoretical expectations, based on the candidate frequency for different braking indices [see Eq. (\ref{Deltaf2}), third block of Fig. \ref{fig:FC}]. The most promising candidates undergo further analysis using the 5-vec Resampling~\cite{sing, Francesco} and BSD Frequency-Hough techniques \cite{FH}, as detailed in Sec. \ref{sec:FU} (fourth block in Fig. \ref{fig:FC}). In the absence of confirmed candidates, we set 95\% confidence-level ULs on the CW strain amplitude shown in Eq. (\ref{strain}) \cite{UL} (fifth block in Fig. \ref{fig:FC}).

\subsection{Tests on simulated data}

\label{sec:bincomb}
Prior to analyzing real data with the search pipeline, we first validated its performance using simulated datasets with CW software injections in Gaussian noise. This validation step was essential to evaluate the effectiveness of the candidate identification strategy in a controlled environment, particularly given the absence of Doppler and spin-down corrections in the analysis, which are not implemented in PyStoch.

In particular, we focused on signals whose expected frequency evolution, due to spin-down effects, would spread across multiple frequency bins [see Eq. (\ref{Deltaf})]. To recover such simulated signals, we applied the frequency bin combination strategy described in Sec.~\ref{SubA}. We used a set of different values of $N$ to enhance the SNR of frequency-varying signals, being aware that the process does not alter the original frequency resolution of the analysis.

A successful test is illustrated in Fig.~\ref{fig:binning}, where a simulated CW signal, injected into Gaussian noise and spread due to spin-down effects, was clearly recovered after applying the appropriate bin combination. The signal, which corresponds to a set of spin-down parameters with $\dot{f}_0 = -10^{-8}$ Hz/s and $\ddot{f}_0 = 10^{-17}$ Hz/s² over a $T_{\text{obs}} = 14$ months observation period, was expected to span 11/32 Hz. With the default resolution, the signal power was spread across several frequency bins, remaining below the detection threshold of SNR$_{\text{thr}} = 4.5$. However, by applying the bin combination strategy with $N=5$, the SNR was effectively enhanced, pushing it above the detection threshold.

\begin{figure}[h!]
\centering
\includegraphics[width=0.50\textwidth]{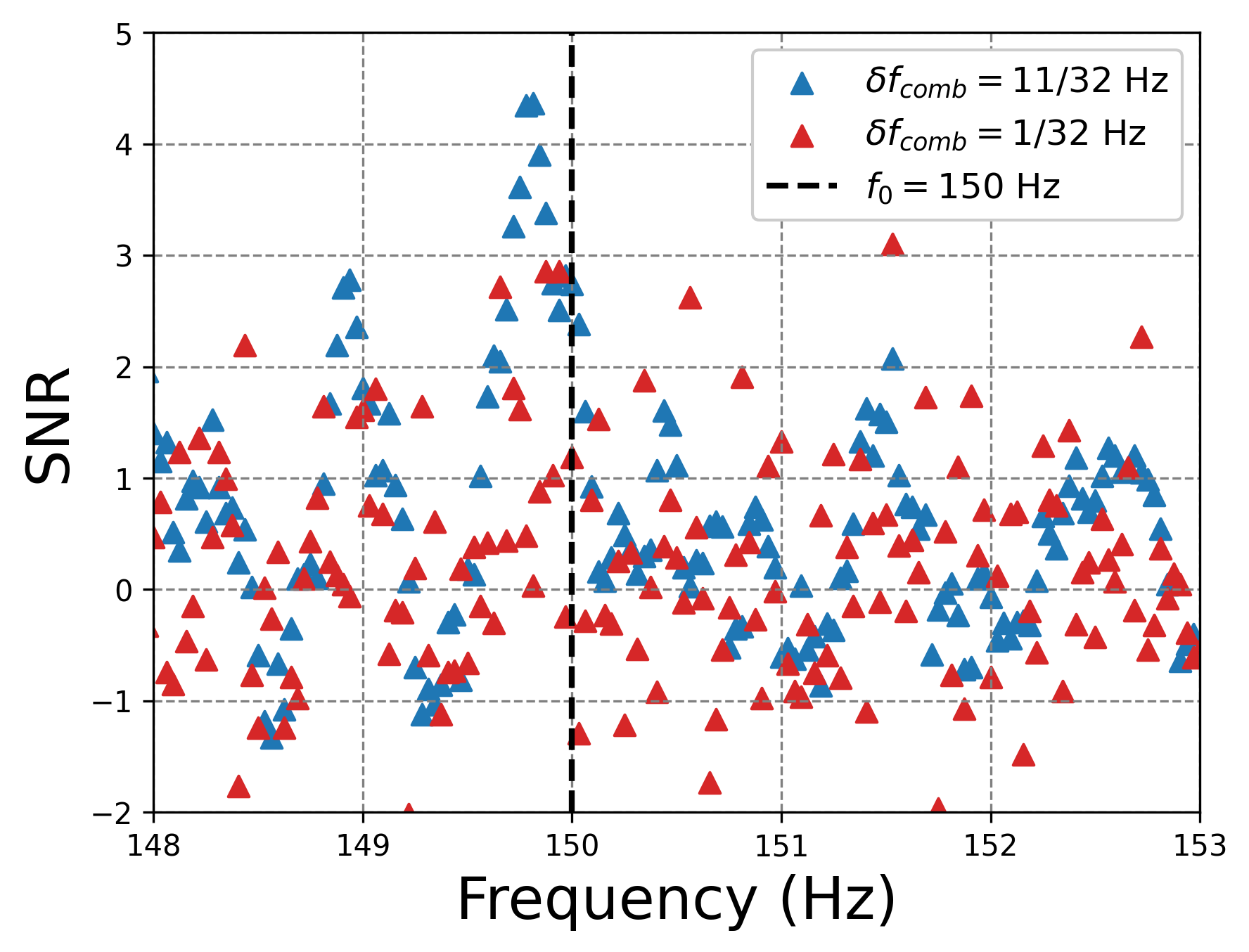}
        \caption[SNR versus frequency after bin combination.]{SNR versus frequency for a dataset processed with PyStoch containing simulated Gaussian noise with an injected CW signal. The dataset corresponds to an observation time $T_{\text{obs}} = 14$ months and a frequency resolution of $\delta f_{\text{def}} = 1/32$ Hz. The red triangles represent the dataset without bin combination (i.e. $N=0$ and $\delta f_{\text{comb}} = \delta f_{\text{def}}$), while the blue triangles correspond to the dataset after the bin combination performed with $N=5$, i.e. $\delta f_{\text{comb}} = 11/32$ Hz. The fake CW signal has $h_0 = 2.2 \times 10^{-25}$, $f_0 = 150$ Hz (dashed line), and is spread over 11 default frequency bins ($\Delta f_0 = 11/32$ Hz) due to its spin-down parameters, i.e. $\dot{f}_0 = -10^{-8} $ Hz/s and $\ddot{f}_0 = 10^{-17}$ Hz/s$^2$.
        }
    \label{fig:binning}
\label{fig:binning}
\end{figure}
\begin{figure}[h!]
\centering
\includegraphics[width=0.50\textwidth]{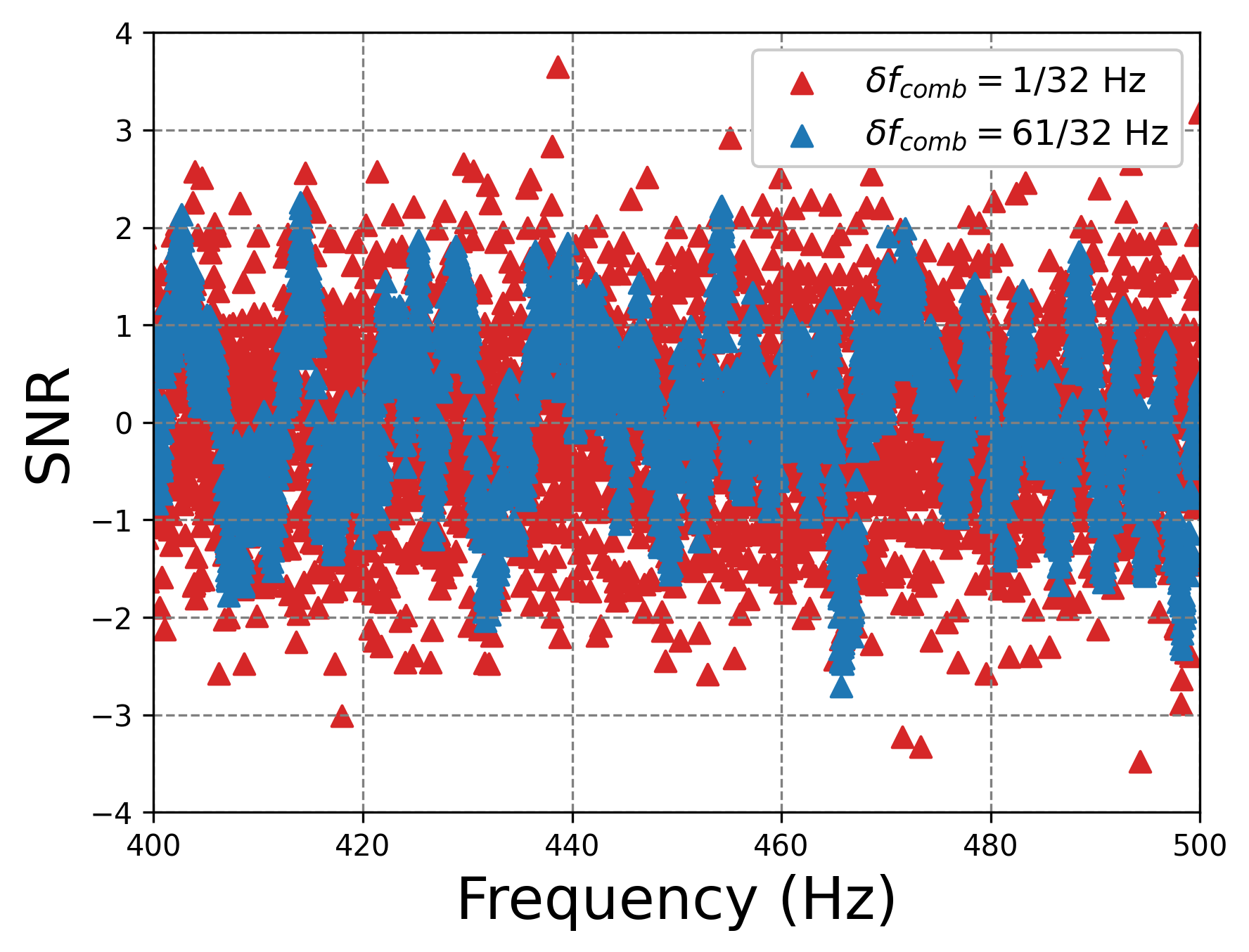}
 \caption{SNR versus frequency for a dataset processed with PyStoch, containing pure simulated Gaussian noise with $T_{\text{obs}} = 14$ months and frequency resolution of $\delta f_{\text{def}} = 1/32$ Hz. The red triangles represent the dataset without bin combination (i.e. $N=0$ and $\delta f_{\text{comb}} = \delta f_{\text{def}}$), while the blue triangles correspond to the dataset after the bin combination performed with $N=30$, i.e. $\delta f_{\text{comb}} = 61/32$ Hz.}
\label{fig:NOISE}
\end{figure}

These tests also provided insight into the practical limits of bin combination. As shown in Fig.~\ref{fig:NOISE}, excessive combination (e.g., $N=30$ or $\delta f_{\text{comb}} = 61/32$ Hz) can lead to noise clustering, creating spurious high-SNR regions that interfere with candidate selection. Through empirical analysis, we determined that frequency combinations exceeding $\delta f_{\text{comb}} = 47/32$ Hz (i.e., $N=23$) significantly degrade the performance, establishing this as the maximum value for effectively enhancing resolution.

\subsection{Selection of candidates}
\label{sec:selection}
The candidate selection process\footnote{Applying the bin-combination strategy by fixing the parameter $N$, we obtain new SNR distributions with the same, fixed bin width of $\Delta f$. Each candidate (with ${\rm SNR} \geq 4.5$) identified at frequency $f_0$ is therefore distributed across a frequency interval of width $\Delta f$. To assess the physical relevance of a candidate, it is necessary to verify that there exists a braking index $n$ within the explored range such that, given the target age $\tau$ and the candidate frequency $f_0$, the predicted frequency evolution as a function of $(f_0, n, \tau)$ matches the observed spread $\Delta f$.} requires that two key conditions are simultaneously satisfied after the bin combination: a statistically significant SNR, i.e. SNR $>4.5$ which corresponds to a $p$ value$<$ 10\%; an observed frequency and frequency distribution of the candidate compatible with theoretical predictions (see Sec. \ref{sec:model}).
Suppose a candidate with SNR $>$ 4.5 was identified in O3 at a frequency $ f_0$ using the bin combination strategy with a specific $\delta f_{\text{comb}}$, i.e. with a specific number of combined bins $N$.
If the same candidate appears with SNR$>4.5$ for multiple values of $N$, only the instance for which the corresponding $\delta f_{\text{comb}}$ is consistent with the expected $\Delta f$ is retained.

Assuming the candidate remains confined within its combined frequency range, then $\delta f_{\text{comb}}$ matches the candidate frequency distribution $\Delta f$ of Eq. (\ref{TOTspread}). Assuming a candidate frequency $f_0$, a braking index $n$, and a source age $\tau$, the candidate is selected if, satisfying the condition SNR$>$4.5, it comes from a bin combination such that $\delta f_{\text{comb}}$ is consistent with the theoretical expectation described in Eq. (\ref{TOTspread}).

The parameter set ($ f_0 $, $n$, $\tau$) defines a specific hypothesis for the candidate spin-down parameters $\dot{f}_0$ and $\ddot{f}_0$ (see Sec. \ref{sec:model}). The $N$-bin combination that triggers a candidate to be selected within this hypothesis is a good criterion to reduce false positives. Furthermore, complementary CW follow-up methods allow to deeply inspect the candidate selection.
In the absence of a detection, ULs are computed (see Sec. \ref{subsec: uls}).

\subsection{Follow-up of candidates}
\label{sec:FU}
When a promising candidate is found, CW detection techniques are used to further investigate and confirm it. Two key methods used for this purpose are the 5-vec Resampling \cite{sing} \cite{Francesco} and the frequency-Hough transform \cite{FH}.

The 5-vec Resampling method begins with the inverse Fourier transform of the frequency domain data back into time series, followed by downsampling and demodulation to correct for Doppler shifts and spin-down effects. The signal power is redistributed across five characteristic frequencies, and when the sky location is known, template-based matched filtering is applied to these peaks, enhancing candidate confirmation \cite{Francesco}.

The BSD frequency-Hough transform is an implementation of the Hough transform pattern recognition algorithm for GW searches. Fixing a sky position, it maps a time-frequency collection of the most significant spectral peaks in the data onto the frequency-spin-down portion of the parameters space, enabling the identification of coherent CW signal traces even in noisy data. By cross-checking events across multiple detectors, this method increases robustness against noise and enhances detection confidence \cite{FH}.

\subsection{Computational cost}
\label{sec:cost}

To better understand the computational advantages of using folding and PyStoch, we can compare the time required by PyStoch and the frequency-Hough transform for searches towards specific sky directions during O3.
The computation time needed by the BSD GPU-frequency-Hough \cite{iuri} to perform a targeted search in O3 with the BSD frequency-Hough transform depends on several factors, including the used device, the frequency band, the number of sky points, and the range of first-order spin-down parameters considered. However, a comparison of the order of magnitude remains highly valuable: searching for CW signals over a range of first-order spin-down parameters between $-10^{-8}$ Hz/s and $+10^{-9}$ Hz/s, from the four targets in Table \ref{tab:targets}, within the [$20-1726$]~Hz frequency band, takes $\sim 6.2$ hours using a single Nvidia V100 \cite{iuri}.

In contrast, performing the same search, still in O3 and with same frequency range, using PyStoch on a CPU with four simultaneous threads, takes a total of approximatively 0.5 hour.

The key trade-off is that, while CW searches have coherence times of  $O(10^3)$ seconds, the radiometer method, constrained by a 192-second coherence time, offers lower sensitivity. This makes the latter, when applied to folded data, an excellent tool for identifying interesting outliers to be followed up with CW methods.


Regarding the 5-vec Resampling \cite{Francesco}, analyzing a 1 Hz-wide band of the full O3 dataset for a single detector using a single-core CPU job takes approximately 8.7 CPU hours (elapsed on a machine with 11.1 HEPSPEC per core). Running an equivalent analysis on a machine equipped with an NVIDIA Quadro P5000 GPU, using a prototype GPU-enabled version of the code, reduces the elapsed time to about 21.3 minutes (0.35 hours). The GPU implementation is still under active development, and preliminary tests suggest that further optimization could yield an overall speed-up of up to a factor of 20 relative to the CPU baseline.

\section{Search in O3 data}
\label{sec:search}
The method described above was applied to the O3 data for the four supernova remnants under investigation, with their known parameters listed in Table \ref{tab:targets}.

\begin{table}[t]
    \centering
\begin{tabular}{lcccc} 
\toprule Target & CasA &Vela Jr. & G347 & SN1987a\\ 
\midrule
 Distance [Kpc] & 3.3 & 0.2-0.9 & 0.9 & 51.4\\ 
 Right ascension [Rad] & 6.124 & 2.321 & 4.509 & 1.464 \\ 
 Declination [Rad] & 1.026 & $-$0.808 & $-$0.695& $-$1.210 \\ 
 Birth year & 1670 & 1300 & 400 & 1987 \\ 
\bottomrule
\end{tabular}
    \caption{Distance, sky position, explosion year and age of CasA, Vela Jr., G347 and SN1987a \cite{panagia}\cite{update}.}
    \label{tab:targets}
\end{table}

Considering Eqs. (\ref{fdot}) and (\ref{Deltaf2}), we assume a braking index $ 5 \leq n \leq 7 $ for SN1987a \cite{owen} and $ 2 \leq n \leq 7 $ for the other targets \cite{update}, with a search frequency range of $ 20 \leq f_0 \leq 1726$ Hz for all targets. The spin-down parameters and the corresponding frequency distribution of hypothetical CW signals in O3 can be computed following Eqs. (\ref{spindown}) to (\ref{Deltaf2}), with results shown in Table \ref{para}.

\begin{table}[b]
    \centering
    \begin{tabular}{cccc}
    \toprule
       Target  & $|\Dot{f}_0| $ [Hz/s] & $\ddot{f}_0$ [Hz/s$^2$] & $|\Delta f_0|$ [Hz] \\
       \midrule
       CasA  & [3.52e-10, 1.82e-7] & [4.34e-20, 3.86e-17] & [0.010, 5.19]  \\ 
       G347  & [6.61e-11, 3.42e-8]  & [1.53e-21, 1.36e-18] & [0.0019, 0.98]  \\ 
       Vela Jr.  & [1.51e-10, 7.82e-8] & [7.98e-21, 7.08e-18] & [0.0043, 2.23]  \\ 
       SN1987a  & [3.20e-9, 4.15e-7] & [3.60e-18, 4.98e-16] & [0.090, 11.66]  \\ 
       \bottomrule
    \end{tabular}
\caption{Expected ranges of $\Dot{f}_0$, $\ddot{f}_0$ and $|\Delta f_0|$ for CasA, G347, Vela Jr., and SN1987a in O3.}
    \label{para}
\end{table}

Given the parameters in Table \ref{para}, the selection method described in Sec. \ref{sec:selection} was applied. As stated in Sec. \ref{sec:selection}, only candidates with SNR $>$ 4.5, and whose frequency and frequency distribution are consistent with theoretical expectations, were selected [based on the target age at the start of O3 and the possible braking index values, see Eq. (\ref{Deltaf2})].\\
We clarify that the follow-up techniques (5-vec Resampling and frequency-Hough) 
were not applied to O3 candidates within the present analysis due to the absence 
of significant outliers. However, both methods have been extensively validated 
on simulated signals and on earlier data, as documented in 
Refs.~\cite{Francesco,FH,sing}. These studies demonstrate the expected 
performance of the follow-up pipeline, ensuring its readiness for future analyses.

\begin{figure}[h]
    \centering
    \includegraphics[width=0.9\linewidth]{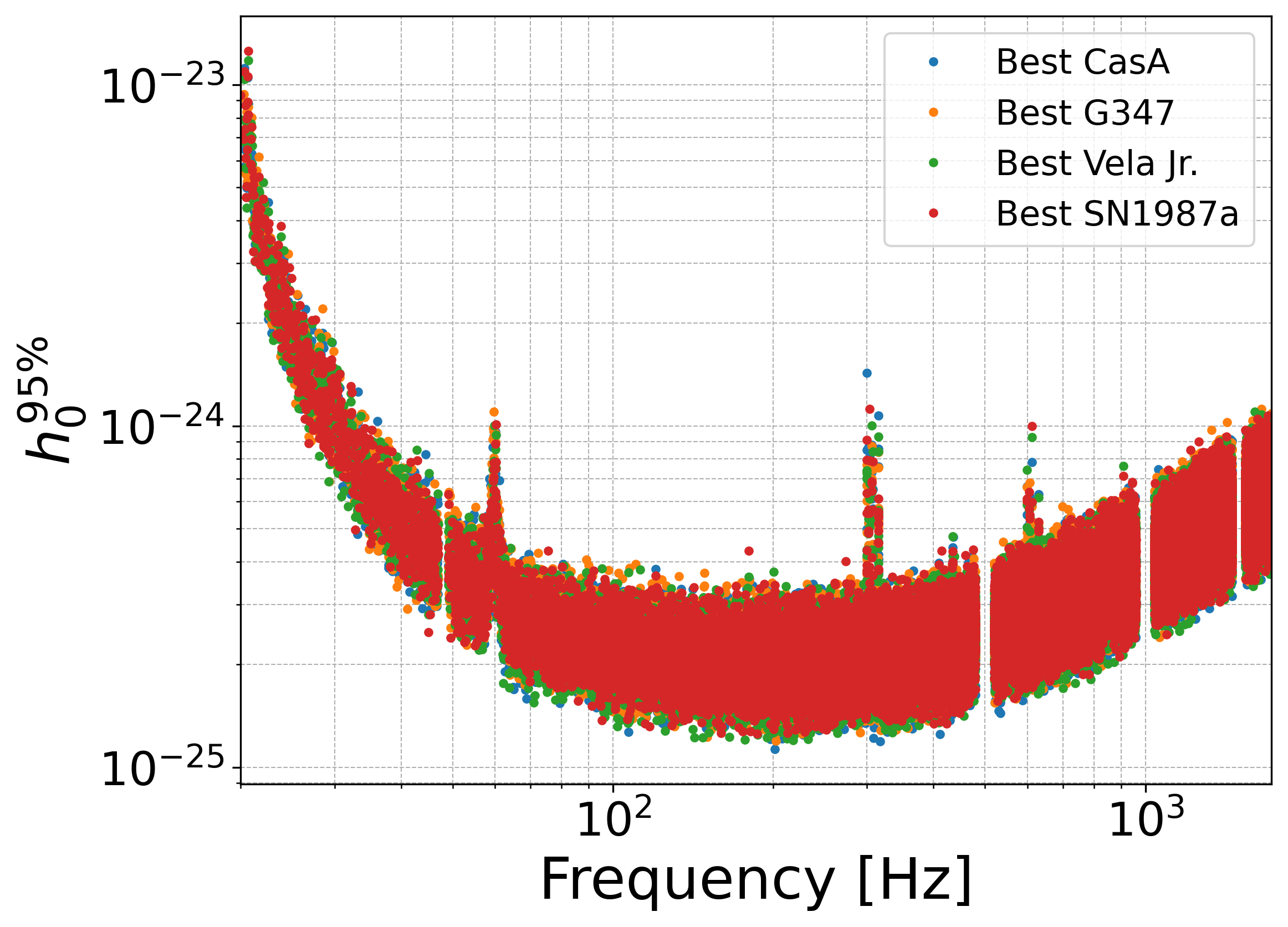}
    \includegraphics[width=0.9\linewidth]{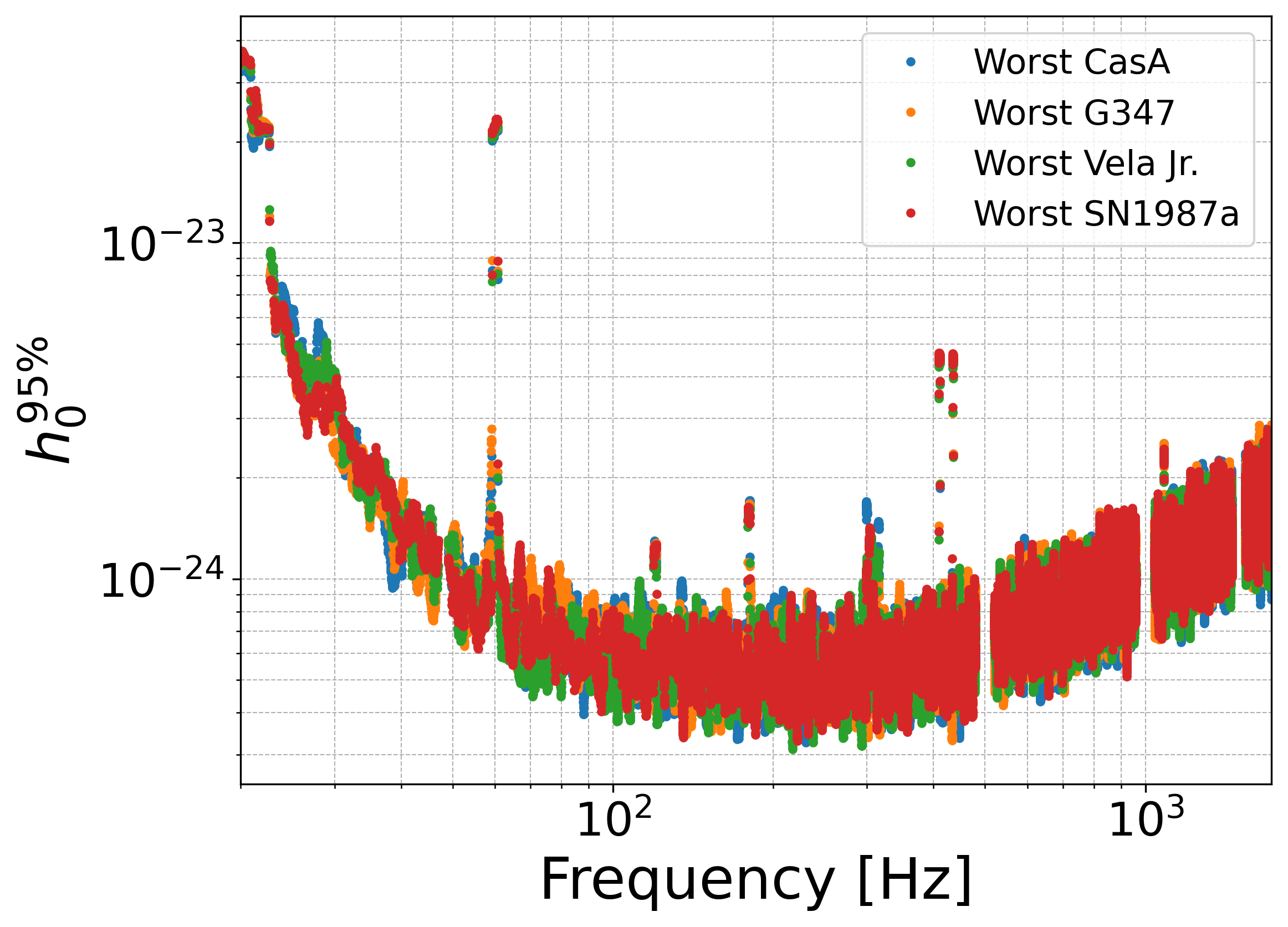}
    \caption{Best ( $\delta f_{\text{comb}}=$ 1/32 Hz, top) and worst ($\delta f_{\text{comb}}=$ 47/32 Hz, bottom) 95\% confidence-level ULs between 20 and 1726 Hz, computed with PyStoch for each target: SN1987a (red), Vela Jr. (green), G347 (orange) and CasA (blue).}
    \label{fig:upperlimits}
\end{figure}

\begin{figure}[h]
    \centering
    \includegraphics[width=1\linewidth]{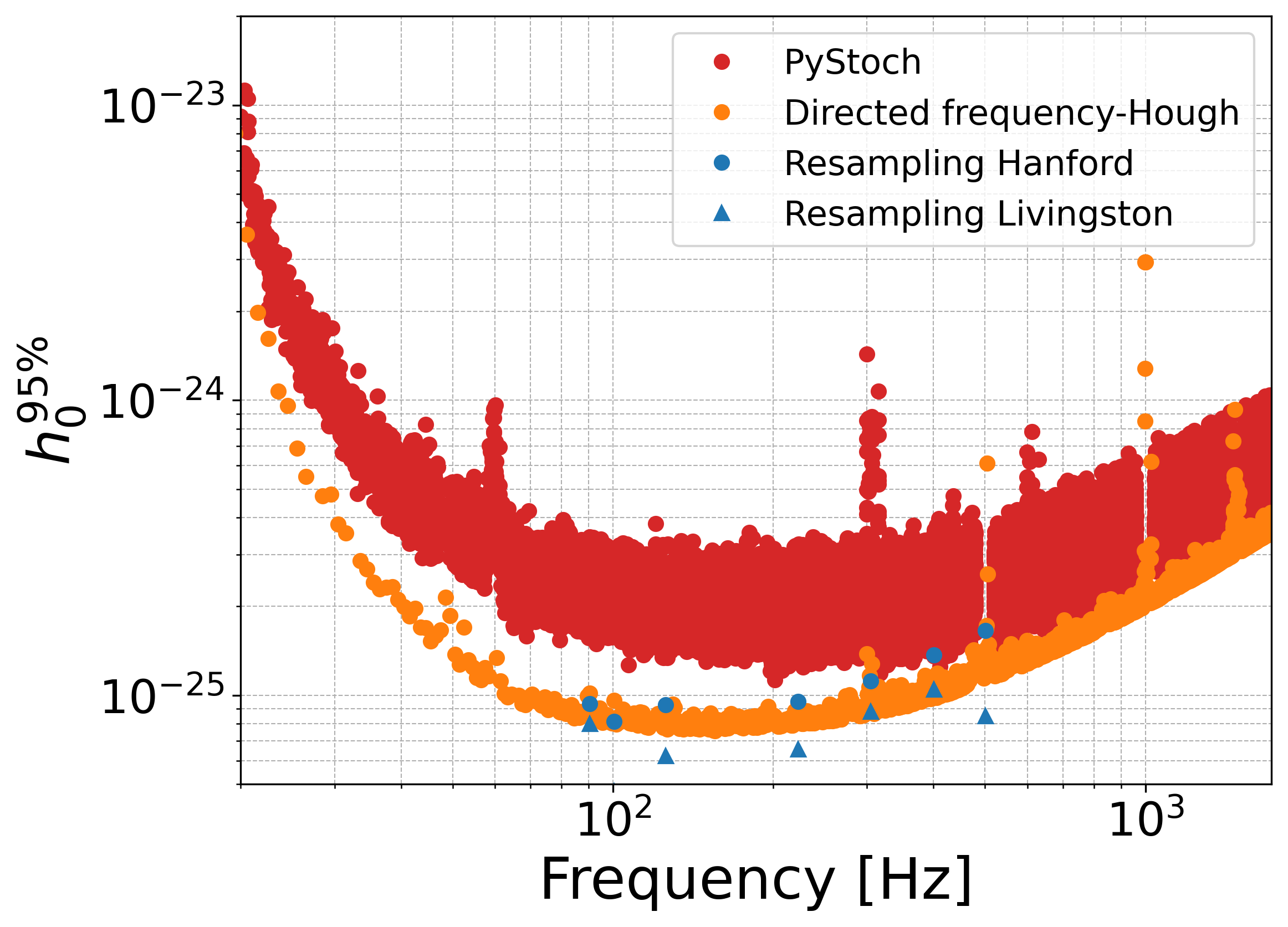}
    \caption{PyStoch best ULs computed targeting CasA (red), frequency-Hough ULs computed targeting the Galactic Center (orange), both between 20 and 1726 Hz, and Resampling ULs computed targeting Scorpius-X1 at selected frequencies for both Livingston (blue triangles) and Hanford (blue circles). 
    We remark that this is an order-of-magnitude comparison of quantities that are not significantly affected by changes in the sky position.}
    \label{fig:dirFH}
\end{figure}

Finally, 95\% confidence-level ULs for the strain amplitude were calculated across varying frequency resolutions, with the best and worst 95\% confidence-level ULs obtained by combining $N=0$ bins (default, $\delta f_{\text{def}} = 1/32$ Hz) and $N=23$ bins ($\delta f_{\text{comb}} = 47/32 $ Hz), respectively, as reported in Table \ref{uls}. In particular, the ULs degrade as $N$ increases.
\begin{table}[h]
    \centering
    \begin{tabular}{lcc}
\toprule
        Target & Best UL $\times 10^{-25}$ & Frequency [Hz] \\
\midrule
        CasA & $1.129 $ & 201.56 \\
        G347 & $1.195 $ & 202.16 \\
        Vela Jr. & $1.198$ & 217.81 \\
        SN1987a & $1.465$ & 186.41 \\
\midrule
        Target & Worst UL $\times 10^{-25}$ & Frequency [Hz] \\
\midrule
        CasA & $3.328 $ & 230.19 \\
        G347 & $3.304 $ & 219.78 \\
        Vela Jr. & $3.115$ & 217.22 \\
        SN1987a & $3.296$ & 221.94 \\
        \bottomrule
    \end{tabular}
    \caption{Best ($\delta f_{\text{comb}} = 1/32$ Hz, top) and worst ($\delta f_{\text{comb}} = 47/32$ Hz, bottom) 95\% confidence-level ULs with corresponding frequency for each target.}
    \label{uls}
\end{table}
In conclusion, the ULs for each frequency in the two cases presented in Table \ref{uls} are shown in Figure \ref{fig:upperlimits}. The top plot displays the ULs between 20 and 1726 Hz in O3 for each target with no bin combination (best ULs), while the bottom plot shows the results for the maximum frequency bin combination (worst ULs).
\\
To provide a broader perspective, in Fig. \ref{fig:dirFH} we show an order-of-magnitude comparison of O3 95\% confidence-level ULs obtained with PyStoch in the best case ($\delta f_{\text{def}}=$ 1/32 Hz, no bin combination, red curve) for CasA in the frequency band of [20, 1726] Hz and the directed frequency-Hough search \cite{dirFH} towards the Galactic Center (orange curve) in the same frequency band, with a spin-down range of [$-10^{-8}$, $10^{-10}$] Hz/s. The 5-vec Resampling \cite{Francesco}  search in the frequency band of [10, 1000]~Hz, targeting Scorpius-X1 is also shown as blue dots and triangles. As expected, PyStoch performs worse in terms of accuracy, but it is exceptionally fast.

\section{Discussion and outlook for future work}
\label{sec:outro}

In this study, we assessed the performance of PyStoch to detect CWs from four notable supernova remnants: Vela Jr., G347, CasA, and the NS associated with SN1987a (see Table \ref{tab:targets}). Using O3 data in the [20–1726]$-$Hz frequency range, we investigated the feasibility of detecting CWs with PyStoch across a range of spin-down parameter combinations (see Table \ref{para}).

Our findings indicate that while stochastic directional searches are computationally efficient, they are less sensitive than traditional CW pipelines (see Secs. \ref{sec:cost} and \ref{sec:search}). To address this limitation, PyStoch can be used in combination with dedicated CW follow-up techniques, which can be employed when significant candidates are identified. 

Since no candidate met the selection criteria outlined in Sec. \ref{sec:selection}, we computed 95\% confidence-level ULs for the CW strain amplitude by combining adjacent frequency bins. The default frequency resolution is $\delta f_{\text{def}} = 1/32$ Hz, while higher effective bin widths are given by $\delta f_{\text{comb}} = (2N+1)\delta f_{\text{def}}$, with $N$ ranging from 1 to 23. These bin widths were obtained by combining adjacent bins through a running average. The most stringent ULs were obtained at the default resolution of 1/32 Hz (e.g when no bin combination was applied), while the least sensitive ones result from $N=23$, i.e. when $\delta f_{\text{comb}}= 47/32 $ Hz (see Table \ref{uls}).

By comparing these ULs with those from CW directed searches in O3, we found that PyStoch is less sensitive than directed searches (see Sec. \ref{sec:search}) but it is exceptionally faster (see Sec. \ref{sec:cost}).

Building on these results, we plan to integrate software injections directly into folded data, enabling direct testing without relying on the full GW radiometer pipeline. This approach will facilitate frequentist estimation of ULs, drastically reducing computational costs from months to hours, and, hopefully, contribute to the detection of CW signals. Ultimately, this will strengthen constraints on NS emission models in future LIGO-Virgo-KAGRA observations.

\section*{Acknowledgments}
This research has made use of data or software obtained from the Gravitational Wave Open Science Center, a service of the LIGO Scientific Collaboration, the Virgo Collaboration, and KAGRA. This material is based upon work supported by NSF's LIGO Laboratory which is a major facility fully funded by the National Science Foundation, as well as the Science and Technology Facilities Council (STFC) of the United Kingdom, the Max-Planck-Society (MPS), and the State of Niedersachsen/Germany for support of the construction of Advanced LIGO and construction and operation of the GEO600 detector. Additional support for Advanced LIGO was provided by the Australian Research Council. Virgo is funded, through the European Gravitational Observatory (EGO), by the French Centre National de Recherche Scientifique (CNRS), the Italian Istituto Nazionale di Fisica Nucleare (INFN) and the Dutch Nikhef, with contributions by institutions from Belgium, Germany, Greece, Hungary, Ireland, Japan, Monaco, Poland, Portugal, Spain. KAGRA is supported by Ministry of Education, Culture, Sports, Science and Technology (MEXT), Japan Society for the Promotion of Science (JSPS) in Japan; National Research Foundation (NRF) and Ministry of Science and ICT (MSIT) in Korea; Academia Sinica (AS) and National Science and Technology Council (NSTC) in Taiwan.
The authors are grateful for computational resources provided by the LIGO Laboratory and supported by National Science Foundation Grants PHY-1626190 and PHY-2110594, as well as for computational resources provided by Virgo's Rome~1 group and supported by INFN.
We would like to express our sincere gratitude to Jishnu Suresh and Deepali Agarwal for their invaluable guidance in familiarizing us with PyStoch, as well as Deepali Agarwal, Bryn Haskell and Karl Wette for their kind assistance in reviewing the work presented here.

\section*{DATA AVAILABILITY}

The data that support the findings of this article are openly available \cite{DataAvail}.